# CT to Cone-beam CT Deformable Registration With Simultaneous Intensity Correction

Xin Zhen[1,2], Xuejun Gu[1], Hao Yan[1], Linghong Zhou[2], Xun Jia[1], and Steve B. Jiang[1]

[1]Center for Advanced Radiotherapy Technologies and Department of Radiation Medicine and Applied Sciences, University of California San Diego, La Jolla, CA 92037-0843, USA
[2]Department of Biomedical Engineering, Southern Medical University, Guangzhou, Guangdong 510515, China

E-mails:  sbjiang@ucsd.edu (Jiang), xunjia@ucsd.edu (Jia), smart@smu.edu.cn (Zhou)

Computed tomography (CT) to cone-beam computed tomography (CBCT) deformable image registration (DIR) is a crucial step in adaptive radiation therapy. Current intensity-based registration algorithms, such as demons, may fail in the context of CT-CBCT DIR because of inconsistent intensities between the two modalities. In this paper, we propose a variant of demons, called Deformation with Intensity Simultaneously Corrected (DISC), to deal with CT-CBCT DIR. DISC distinguishes itself from the original demons algorithm by performing an adaptive intensity correction step on the CBCT image at every iteration step of the demons registration. Specifically, the intensity correction of a voxel in CBCT is achieved by matching the first and the second moments of the voxel intensities inside a patch around the voxel with those on the CT image. It is expected that such a strategy can remove artifacts in the CBCT image, as well as ensuring the intensity consistency between the two modalities. DISC is implemented on computer graphics processing units (GPUs) in compute unified device architecture (CUDA) programming environment. The performance of DISC is evaluated on a simulated patient case and six clinical head-and-neck cancer patient data. It is found that DISC is robust against the CBCT artifacts and intensity inconsistency and significantly improves the registration accuracy when compared with the original demons.

**1. Introduction**

Online Adaptive Radiation Therapy (ART) allows real-time treatment adaptations based on the current patient anatomy and geometry. In a typical online ART process, a computed tomography (CT) image is usually acquired prior to the treatment course for treatment planning purposes. Before each treatment fraction, a cone-beam computed tomography (CBCT) image is then obtained, on which the treatment plan is redesigned to account for setup errors, deformations of tumor and other organs, as well as the change of their relative locations. Deformable image registration (DIR) technique plays an important role in this process to establish a correspondence between voxels in the CT and the CBCT for various purposes, for instance, transferring the organ contours from the planning CT images to the daily CBCT images. It is hence desirable to have an accurate and robust DIR algorithm to facilitate this step.

Among the existing DIR methods, demons (Thirion, 1998) has been proven to be a fast and robust algorithm and a number of its variants have been developed (Pennec *et al.*, 1999; Wang *et al.*, 2005; Rogelj and Kovacic, 2006; Yang *et al.*, 2008). However, the demons algorithm assumes that there exists intensity consistency between two images to be registered. Therefore, although demons can successfully deal with images of the same modality (e.g., CT-CT registration), considerable registration error may present, when it comes to inter-modality registration problems (e.g., CT-MRI registration), where corresponding points in the two images to be registered do not necessarily pertain to the same intensity level (Guimond *et al.*, 2001; Lu *et al.*, 2010; Nithiananthan *et al.*, 2011; Hou *et al.*, 2011).

CT-CBCT DIR is considered to be an inter-modality DIR problem. Although CT and CBCT are reconstructed under the same physical principles, the intensity (Hounsfield Units (HU)) consistency between CT and CBCT images is violated due to many reasons. Firstly, almost all current commercial systems reconstruct CBCT images using the well-known FDK (Feldkamp *et al.*, 1984) algorithm. As a fundamental limitation of this algorithm, CBCT quality degrades with increasing of cone angle (Schulze *et al.*, 2011). Second, scatter contamination also leads to severe cupping, streak artifacts and degrade of image contrast. In a typical clinical CBCT system for radiotherapy scatter-to-primary ratio (SPR) may even exceed 100% (Siewerdsen and Jaffray, 2001), although many methods have been proposed to correct scatter artifacts (Siewerdsen *et al.*, 2006; Rinkel *et al.*, 2007; Maltz *et al.*, 2008; Zhu *et al.*, 2009b; Poludniowski *et al.*, 2009; Yan *et al.*, 2010; Meyer *et al.*, 2010; Sun *et al.*, 2011), it is still an open problem. Third, the gantry mounted bowtie filter in CBCT system may wobble, as the gantry rotates, which can result in crescent artifacts (Giles *et al.*, 2011; Zheng *et al.*, 2011). Moreover, there are also other factors that contribute to the intensity inconsistency between CT and CBCT, e.g. different level of noise, beam hardening effects and motion (Zhu *et al.*, 2009a; Hsieh *et al.*, 2000; Grimmer and Kachelriess, 2011; Li *et al.*, 2006; Lewis *et al.*, 2011) .

Despite the difficulties caused by the different image intensities, it is still desirable to use demons-type algorithm in DIR due to its simplicity and hence high efficiency and robustness(Sharp *et al.*, 2007; Gu *et al.*, 2010). As such, a variety of inter-modality demons methods have been proposed. Some researchers incorporate more reliable statistical similarity metrics into demons, such as normalized cross-correlation (NCC) or normalized mutual information (NMI), to measure the similarity between corresponding anatomical points. For instance, Modat *et al* (2010) implemented a diffeomorphic demons using the analytical gradient of NMI in a conjugate gradient optimizer. Lu *et al* (2010) has also proposed a variational approach for multimodal image registration based on the diffeomorphic demons algorithm by replacing the standard demons similarity metric with point-wise mutual information in the energy function. However, it is still unclear what alternative metrics is robust for this CT-CBCT DIR problem and how to incorporate it into demons style algorithms. On the other hand, some researchers focused on estimating the intensity relation between CT and CBCT images and combining intensity correction with geometrical transformation. Within this category, Guimond *et al* (2001) investigated the functional transformation that maps the intensities of one image to those of another, and implemented the intensity correction prior to each iteration. However, the intensity mapping of a polynomial form estimated globally based on the entire image may not be accurate enough and hence may result in errors in subsequent registration process. Recently, Hou *et al* (2011) attempted to correct image intensities by aligning the cumulative histograms of the two images. Nithiananthan *et al* (2011) also tried to correct intensities at each demons iteration by estimating linear transformations for intensity mapping for a few segmented tissue types. Nevertheless, all these works essentially assume that there exists a global mapping between the CT and the CBCT intensities, which, however, may not hold when CBCT is contaminated by artifacts, whose pattern is usually local.

In this work, we propose and evaluate a modified demons algorithm embedded with a simultaneous intensity correction step, called *Deformation with Intensity Simultaneously Corrected* (DISC). Rather than estimating a global mathematical transformation model between CT and CBCT images, our method corrects CBCT intensity of each voxel at every iteration step of demons by matching the first and the second moments of the voxel intensities inside a patch around this voxel with those in the CT image. Quantitative evaluations of our method are performed by using both a simulation data set and six clinical head-and-neck cancer patient data sets. It is found that DISC can handle CBCT artifacts and the intensity inconsistency issue and therefore improves the registration accuracy when compared with the original demons.

## 2. Methods and materials

*2.1 The original demons algorithm*

Suppose we would like to register two images $I_m(x)$ and $I_s(x)$, where a vector field $v(x)$ relates these two images by $I_m(x + v(x)) = I_s(x)$. Because $I_m$ is gradually deformed to match $I_s$, it is termed as the *moving image*, while $I_s$ is called the *static image*. In the demons algorithm, the vector field $v(x)$ is solved in an iterative fashion and at each iteration the increment of the vector field $d\mathbf{r}(x)$ is determined based on the image intensity at the voxel $x$. There are six different variants of the demons algorithm that have been studied by Gu et al (2010). The difference between them lies in the expression for computing $d\mathbf{r}(x)$. Take the double fource demons (Wang *et al.*, 2005; Rogelj and Kovacic, 2006) as an example, the demons algorithm iteratively performs the following steps. First, calculate the increment of the moving vector (or the displacement vector) at all voxel points. Specifically, $d\mathbf{r}=(dx,dy,dz)$ at a voxel in the double force demons is:

$$d\mathbf{r}^{(k+1)} = \frac{(I_m^{(k)} - I_s)\nabla I_s}{(I_m^{(k)} - I_s)^2 + |\nabla I_s|^2} + \frac{(I_m^{(k)} - I_s)\nabla I_m^{(k)}}{(I_m^{(k)} - I_s)^2 + \left|\nabla I_m^{(k)}\right|^2}, \quad (1)$$

where the superscript indexes the iteration step, $I_m^{(k)}$ is the intensity of the moving image at the *k*th iteration, $I_s$ is the original static image. Second, smooth the resulting incremental vector field $d\mathbf{r}$ by convolving it with a Gaussian kernel. Third, add the incremental deformation field to the global deformation field $v(x)$ and update the moving image. This process is iteratively performed until convergence.

*2.2 The DISC algorithm*

In the context of CT-CBCT registration, the CT image is used as the moving one while the CBCT image is the static one. Because of the relative better image quality of CT, deforming CT yields better-controlled numerical error than deforming the CBCT. Due to the intensity inconsistency between CT and CBCT, the original demons algorithm usually fails in CT-CBCT DIR. The DISC algorithm presented in this paper solves this problem by integrating a novel local intensity correction step into the demons framework. The rationale behind DISC is that, although it is generally difficult to establish a global intensity mapping between CT and CBCT, there exists such a mapping locally at each voxel to convert the intensity from CBCT to CT. As such, we insert a CBCT intensity correction step before the first step of the original demons algorithm. This step estimates an intensity transformation at a voxel $x$ based on voxels in small cubic volumes centering at $x$ in both CT and CBCT (termed *patches*). Then the intensity of the CBCT image is adjusted at each voxel using the estimated voxel-dependent transformations. By performing the estimation locally on small patches around each voxel, one can effectively remove image artifacts.

Suppose at an intermediate step $k$ of the registration process, the moving image is deformed into $I_m^{(k)}$. We would like to estimate two parameters $a$ and $b$ at each voxel $x$, such that the CBCT intensity $I_s(x)$ is corrected into $a(x)I_s(x) + b(x)$. This is achieved by comparing the patches centering at $x$ on $I_m^{(k)}$ and $I_s$. Let us consider two patches of size $n$: $\mathbf{M}\{m_1, m_2, \cdots, m_n\}$ and $\mathbf{S}\{s_1, s_2, \cdots, s_n\}$, both centered at voxel $x$ in $I_m^{(k)}$ and $I_s$, respectively. It is our objective to find a linear intensity mapping:

$$\mathbf{S}' = a(x)\mathbf{S} + b(x), \quad (2)$$

such that the intensity distributions of $\mathbf{S}'$ are in agreement with $\mathbf{M}$. This can be achieved by matching the moments of intensity distributions of the two groups of voxels. The *p*-th order raw moment (also known as sample moment) of patch $\mathbf{X}\{x_1, x_2, \cdots, x_n\}$ is defined as:

$$\mathrm{E}(\mathbf{X}^p) = \frac{1}{n}\sum_{i=1}^{n} x_i^p. \tag{3}$$

In this work, only the first two moments are used:

$$\mathbf{S}' = a(x)\mathbf{S} + b(x),$$

s.t.

$$\mathrm{E}(\mathbf{M}) = \mathrm{E}(\mathbf{S}'),$$
$$\mathrm{E}(\mathbf{M}^2) = \mathrm{E}(\mathbf{S}'^2), \tag{4}$$

where $a(x)$ and $b(x)$ for the voxel $x$ can be obtained by solving the above equations (see Appendix) as:

$$a(x) = \mathrm{STD}(\mathbf{M})/\mathrm{STD}(\mathbf{S}), \tag{5}$$
$$b(x) = \mathrm{E}(\mathbf{M}) - a(x)\mathrm{E}(\mathbf{S}), \tag{6}$$

where $\mathrm{STD}(\cdot)$ is the standard deviation operator. Let $\mathbf{A}$ and $\mathbf{B}$ denote two vectors of length $N$ (number of voxels in $\boldsymbol{I}_m$ or $\boldsymbol{I}_s$) with entries $a$ and $b$, respectively. Strictly speaking, the estimation of $a(x)$ and $b(x)$ in Equations (5) and (6) are valid only when the two voxels $\boldsymbol{I}_m^{(k)}(x)$ and $\boldsymbol{I}_s(x)$ are at the same anatomical location. Yet, due to the apparent violation caused by the image deformation, this estimation is not always reliable. In practice, we only estimate those $a$ and $b$ when $\boldsymbol{I}_m^{(k)}(x)$ and $\boldsymbol{I}_s(x)$ belong to the same tissue class and limit $a$ and $b$ in a certain range to avoid false correction. Specifically, we first define a mask:

$$\xi(x) = \begin{cases} 1, & \text{where } \left|\boldsymbol{I}_m^{(k)}(x) - \boldsymbol{I}_s(x)\right| > \mathrm{HU}_0 \\ 1, & \text{where } a(x) < a_{min} \text{ or } a(x) > a_{max} \\ 0, & \text{elsewhere} \end{cases}, \tag{7}$$

where $\mathrm{HU}_0$ is the threshold in Hounsfield Unit to identify if two voxels in CT and CBCT images belong to the same tissue class. $a_{min}$ and $a_{max}$ are the lower and upper bounds of $a$. The way of choosing the parameters $\mathrm{HU}_0$, $a_{min}$ and $a_{max}$ will be discussed in Section 2.3. For voxels with $\xi = 1$, the values of $a$ and $b$ cannot be reliably calculated using Equations (5) and (6); instead, they are estimated by interpolating/extrapolating from the $a$ and $b$ values for voxels with $\xi = 0$. Initially, we set $a(x) = 0$ and $b(x) = 0$ for $\xi(x) = 1$. We then estimate $a(x)$ and $b(x)$ values for voxels with $\xi = 1$ by computing a weighted average of all $a$ and $b$ values of voxels inside a small cubic region $T$ centering at $x$:

$$a(x) = \sum_{i \in T} \omega_i a_i / \sum_{i \in T} \omega_i, \tag{8}$$
$$b(x) = \sum_{i \in T} \omega_i b_i / \sum_{i \in T} \omega_i, \tag{9}$$
$$\omega_i = \exp\left[-\left(\boldsymbol{I}_s(i) - \boldsymbol{I}_s(x)\right)^2 / h^2\right], \tag{10}$$

where $\omega_i$ is the weighting factor determined by the intensity values of $\boldsymbol{I}_s(i)$. The underlying assumption is that voxels of similar intensities should have similar $a$ and $b$ values. $h$ is a parameter that adjusts to what extent we would like to enforce the similarity. This interpolation/extrapolation step may need to be performed for multiple times since for some voxels, the values of $\xi$ may be 1 for all voxels in their neighbor $T$. We need to point out that this interpolation/extrapolation procedure does not change the values of $a$ and $b$ for for voxels with $\xi = 0$.

As soon as $a$ and $b$ are available at all voxels, an intensity correction is performed to update the image intensity of the original static image $I_s(x)$ to yield an intensity corrected static image $I'_s(x) = a(x)I_s(x) + b(x)$, which is thus ready for the displacement calculation using the original demons algorithm. We would like to point out that any variants of the original demons can be implemented in DISC in a similar way.

*2.3 Implementation of DISC*

Before starting the DIR procedure, a global intensity transformation is first performed to shift the CBCT intensity by a constant (denoted as ΔHU) to match the mean intensities of CT and CBCT. This is to correct the average HU difference between the two images to a certain extent.

Then, a multi-scale strategy is adopted so as to reduce the magnitude of the displacement with respect to voxel size. The iteration starts with the lowest resolution images, and the moving vectors obtained at a coarser level are up-sampled to serve as initial solution at a finer level. In this study, we considered two different resolution levels. Further down-sampling was found not to improve registration accuracy nor efficiency.

Before the moving vector field calculation at each iteration, **A** and **B** are estimated using Equations (4)~(10). Some parameters need to be first determined. We notice that $HU_0$ is a parameter affected by the degree of intensity inconsistency and complexity of artifacts, and therefore is proportional to ΔHU. In this study, ΔHU is about 100 HU for simulation data, and about 100~300 HU for clinical data. We empirically choose $HU_0 = 2\Delta HU$. For $a_{min}$ and $a_{max}$, we can learn from Equation (5) that $a$ is actually the standard deviation ratio of the two patches. Thus, **A** is calculated before starting DIR and the median value $a_m$ of **A** is used to determine the range of $a$ in the subsequent steps. We let $a_{min} = 0.5a_m$ and $a_{max} = 1.5a_m$. As for the size of $T$ in the interpolation/extrapolation procedure to get the masked $a(x)$ and $b(x)$, we choose a relative small size 7×7×3 to balance the efficiency and accuracy.

One of the most commonly used stopping criterion to judge whether the moving image has been correctly deformed to the static image is the cross correlation coefficient (Wang *et al.*, 2005; Sharp *et al.*, 2007; Yang *et al.*, 2008; Samant *et al.*, 2008). However, such a similarity metric does not work well in this CT-CBCT DIR context. We therefore use a convergence criterion based on the difference between successive deformation fields. We define a relative norm $l^{(k)} = \sum |d\mathbf{r}^{(k+1)}| / \sum |\mathbf{r}^{(k)}|$, and use $l^{(k-10)} - l^{(k)} \leq \varepsilon$, where $\varepsilon = 1.0 \times 10^{-4}$ as our stopping criterion. This measure is found to have a closer correspondence with spatial accuracy than correlation coefficient as DIR is stopped when there is no 'force' to push voxels any more (Gu *et al.*, 2010).

In this work, we use the compute unified device architecture (CUDA) architecture with an NVIDIA GPU card as the implementation platform. In order to efficiently parallelize DISC in the CUDA environment, the data parallel portions of the algorithm are identified and grouped into the following kernels: 1) an intensity correction kernel to compute and interpolate **A** and **B**; 2) a Gaussian filter kernel to smooth images and moving vectors; 3) a gradient kernel to calculate the gradient of images; 4) a moving vector kernel to calculate and update moving vectors; 5) an interpolation kernel to deform images with moving vectors; and 6) a comparison kernel to stop the program based on the stopping criteria.

Considering all the components mentioned above, we summarize the DISC algorithm in algorithm A1:

**Algorithm A1:**

Globally adjust the intensity of CBCT

Initialize the moving vector **r** to zero

Down-sample the images to the coarsest resolution

**Repeat for each resolution level**

  **while** $l^{(k-10)} - l^{(k)} \leq \varepsilon$, **do**

  1. Compute $a$ and $b$ for each voxel using patches in $I_m^{(k)}$ and $I_s$ (Eqs. (5), (6));
  2. Interpolate/extrapolate $a$ and $b$ to where $\xi = 1$;
  3. Obtain $I'_s$ by applying the estimated linear transformation at each voxel;
  4. Compute d**r** for each voxel (Eq. (1));
  5. Add d**r** to the total moving vector **r**;
  6. Regularize **r** by applying a Gaussian kernel;

  Up-sample the moving vector **r** to a finer resolution level

**Until** the finest resolution is reached

*2.4 Evaluation*

*2.4.1 Synthetic data: MC simulation*

To validate our algorithm, we have generated a test dataset based on two CT images of a head-and-neck cancer patient. The first CT image is called planning CT acquired before the treatment while the second CT image was acquired half way in the treatment course for re-planning purpose and is called treatment CT here. Then, a CBCT image with realistic image artifacts is synthesized using the treatment CT image. DISC is applied to perform DIR between the planning CT image and the synthesized CBCT image and to correct the CBCT intensity. This approach offers us the ground truth for the evaluation of DISC: the treatment CT can be regarded as the scatter-free CBCT image and the deformation vector field between the planning and treatment CT images obtained using the original demons algorithm is the ground truth deformation vector field.

To synthesize a realistic CBCT image using the treatment CT image, we first convert the CT image into a digital phantom by assigning each voxel with a density value and a material type. CBCT projection images at 360 equally spaced directions covering an entire $2\pi$ angular range are then calculated using an in-house developed software tool (called gDRR) (Jia *et al.*, 2012) under a realistic projection geometry for a Varian On-board-Imaging system (OBI) (Varian Medical Systems, Inc., Palo Alto, CA). In this package, the primary component in a projection image is calculated by a ray-tracing algorithm, while the scatter component is obtained by Monte Carlo simulations followed by an image smoothing process to suppress noise. Both the primary and the scatter calculations consider a variety of effects occurred in a realistic CBCT scan including the energy spectrum, the source fluence map, and the detector response, etc.. Once the projections are generated, an FDK reconstruction algorithm is invoked, yielding the CBCT image with exactly the same anatomy structures as in the treatment CT image but with all major CBCT artifacts such as scatter.

*2.4.2 Clinical data*

The performance of DISC is further assessed using clinical CT and CBCT data of six head-and-neck cancer patients. Each patient has a planning CT image and a CBCT image. The CBCT images were acquired 1-7 weeks after the first fraction of treatment on a Varian OBI system integrated in a Trilogy$^{TM}$ linear accelerator (Varian Medical Systems, Inc., Palo Alto, CA) using full-fan mode with a full-fan bow-tie filter on site.

For the planning CT images, the image resolution in the transverse plane is 512×512 and the slice thickness is either 1.25 or 2.5 mm. The pixel size in the transverse plane varies from 0.74 to 1.07 mm. For all the CBCT images, the image size in the transverse plane is 512×512 and the slice thickness is 2.5mm. The pixel size in the transverse plane is 0.47mm.

The number of transversal slices ranges from 140 to 220 for a CT image and is approximately 70 for a CBCT image. Therefore, the field of view of the planning CT is generally larger than CBCT. The planning CT image is then cropped and re-sampled to match the dimension and resolution of the CBCT image after rigid registration. Both CT and CBCT images are down-sampled to half of their original size in the transverse plane. Hence, the image resolution for both CT and CBCT images after rigid registration is 256×256×68 (Cases 1, 2, 5, 6) or 256×256×52 (Cases 3, 4), and the voxel size is 0.94×0.94×2.5mm.

*2.4.3 Quantification of registration performance*

Three similarity metrics are used in this work to quantify the DIR results. They are chosen based on two considerations. First, the metric should be observer independent. Second, the metric should be insensitive to intensity inconsistency.

The first metric is normalized mutual information (NMI), ranging from 0 to 1 with 1 representing the highest image similarity. The second metric is called feature similarity index (FSIM) (Zhang *et al.*, 2011; Yan *et al.*, 2012), which tries to model the mechanism of the human visual system by capturing the main image features such as the phase congruency of the local structure and the image gradient magnitude. Detailed definition and description of FSIM are given in (Zhang *et al.*, 2011). In this work, FSIM is calculated at each pair of corresponding transverse slices between two 3D data sets, and the average value and standard deviation are calculated. The FSIM score varies between 0 and 1 with 1 representing the most image similarity.

The third metric is the root mean squared error (RMSE) between two edge images:

$$\text{RMSE}_{\text{edge}} = \sqrt{\Sigma_i^N (I_1^{edge}(i) - I_2^{edge}(i))^2 / N}, \quad (11)$$

where $I_1^{edge}(i)$ and $I_2^{edge}(i)$ are the binary Canny edge images of image $I_1$ and $I_2$, respectively (Canny, 1986). When two images are perfectly aligned, $\text{RMSE}_{\text{edge}}$ should be zero.

## 3. Results

For clarity, following symbols are used to represent different images used in the algorithm evaluation: $\text{CT}_{\text{original}}$ and $\text{CBCT}_{\text{original}}$ are the CT and CBCT images before registration, respectively. $\text{CT}_{\text{deformed}}^{\text{demons}}$ and $\text{CT}_{\text{deformed}}^{\text{DISC}}$ are the deformed CT images using the original demons algorithm and the DISC algorithm, respectively. $\text{CBCT}_{\text{corrected}}$ is the intensity corrected CBCT image using DISC. In the simulation study, $\text{CT}_{\text{original}}$ refers to the planning CT before registration; $\text{CBCT}_{\text{original}}$ is the synthesized CBCT using the treatment CT before registration. $\text{CBCT}_{\text{original}}^{\text{primary}}$ is the treatment CT and regarded as the primary part of $\text{CBCT}_{\text{original}}$ before registration; $\text{CT}_{\text{deformed}}^{\text{demons}}(\rightarrow \text{CBCT}_{\text{original}}^{\text{primary}})$ and $\text{CT}_{\text{deformed}}^{\text{demons}}(\rightarrow \text{CBCT}_{\text{original}})$ are the deformed CT images to match $\text{CBCT}_{\text{original}}^{\text{primary}}$ and $\text{CBCT}_{\text{original}}$ using the original demons algorithm, respectively. $\text{CT}_{\text{deformed}}^{\text{DISC}}(\rightarrow \text{CBCT}_{\text{original}})$ is the deformed CT image to match $\text{CBCT}_{\text{original}}$ using DISC.

*3.1 Synthetic data*

Figure 1 shows the results for the simulation case. Figure 1(a) is the planning CT image before registration; Figure 1(b) is the synthesized CBCT image before registration, in which intensity variations due to scatter artifacts can be clearly observed inside the yellow dashed circle. Figure 1(c) is the scatter-free primary component of $\text{CBCT}_{\text{original}}$ before registration

Registration between $\text{CT}_{\text{original}}$ and $\text{CBCT}_{\text{original}}^{\text{primary}}$ (Figure 1(a) and Figure 1(c)) using the original demons algorithm can be regarded as the ground truth which yields $\text{CT}_{\text{deformed}}^{\text{demons}}(\rightarrow \text{CBCT}_{\text{original}}^{\text{primary}})$ (Figure 1(d)). When the original demons algorithm is applied to deform the original CT to the original CBCT, we can see that in $\text{CT}_{\text{deformed}}^{\text{demons}}(\rightarrow \text{CBCT}_{\text{original}})$ (Figure 1(e)) soft tissue and bone inside the yellow dashed circle are significantly distorted after registration due to the intensity inconsistency mainly caused by scatter artifacts. In contrast, DISC can yield correct result, as shown in Figure 1(f).

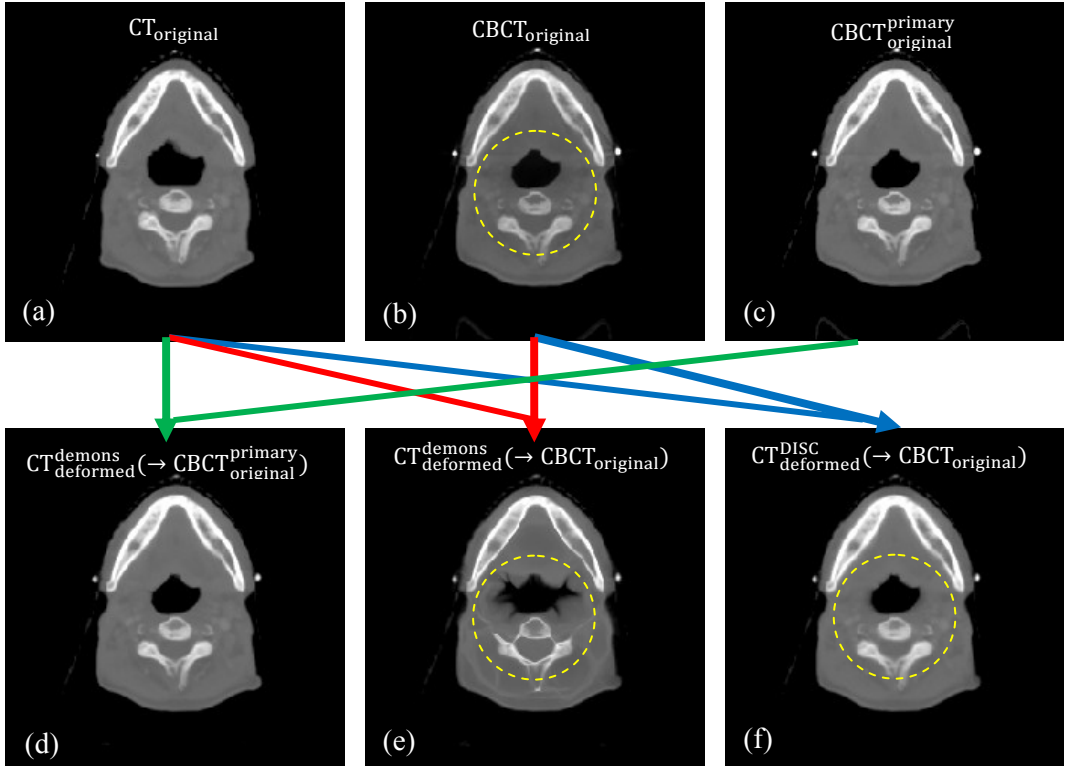

**Figure 1.** Simulation and registration results. The yellow dashed circles indicate the regions which severely suffer from the scatter artifact. The green arrow indicates the DIR between $CT_{original}$ and $CBCT_{original}^{primary}$ using the original demons algorithm; the red arrow indicates the DIR between $CT_{original}$ and $CBCT_{original}$ using the original demons algorithm; the blue arrow indicates the DIR between $CT_{original}$ and $CBCT_{original}$ using DISC.

Checkerboard comparisons are shown in Figure 2. Misalignment is evident before registration (Figures 2(a)-1~3). The original demons algorithm unrealistically distorts the tissues after registration, especially in those central regions with severe scatter artifacts (Figures 2(b)-1~3). Results of DISC are shown in Figures 2(c)-1~3 and Figures 2(d)-1~3. As we can see, DISC is robust against the artifacts and able to match the two images well without distortion (Figure 2(c)-1~3). Checkerboard boundary between $CBCT_{corrected}$ and $CT_{deformed}^{DISC}(\rightarrow CBCT_{original})$ can hardly be seen since $CBCT_{corrected}$ has similar intensity distribution as the deformed CT (Figure 2(d) -1~3).

The improvement of DISC over original demons can be further examined by inspecting the difference of the moving vector fields (Figure 3). Because of the absence of artifacts in $CBCT_{original}^{primary}$, it is expected that the original demons algorithm is functional when it is used for DIR between $CT_{original}$ and $CBCT_{original}^{primary}$, and the resulting deformation vector field $\mathbf{r_0}$ can be regarded as the ground truth vector field. We further denote the vector field obtained by the original demons algorithm and DISC between $CT_{original}$ and $CBCT_{original}$ by $\mathbf{r_1}$ and $\mathbf{r_2}$. The errors in terms of vector field in these two cases are hence characterized by the difference between $|\mathbf{r_1}-\mathbf{r_0}|$ and $|\mathbf{r_2}-\mathbf{r_0}|$, which are shown in Figure 3. We can see that the original demons produces incorrect deformation

vector field when artifacts exist (Figure 3(a)), while the DISC algorithm can yield almost the same deformation vector field as the ground truth (Figure 3(b)).

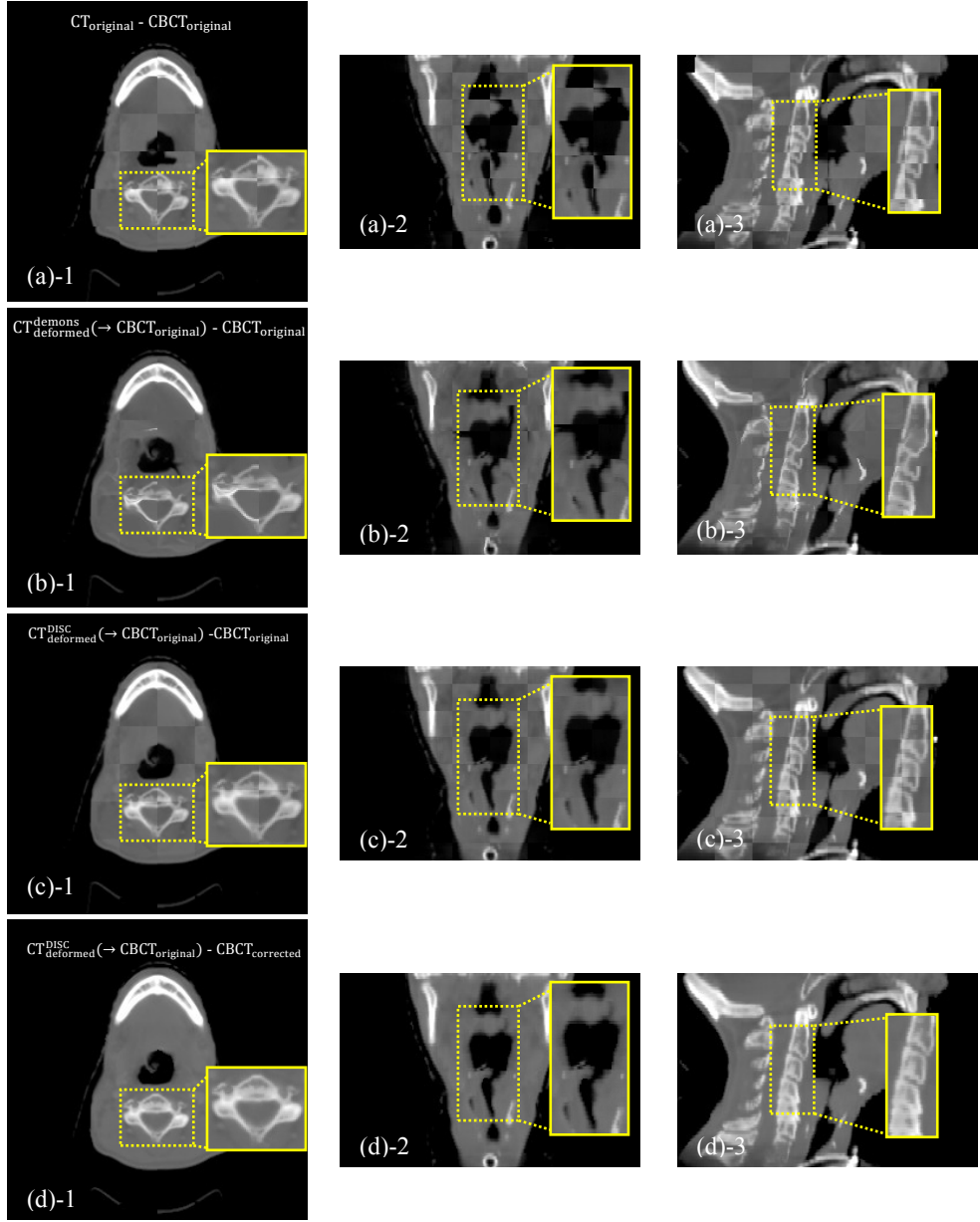

**Figure 2.** Checkerboard comparisons of the DIR results based on the simulation dataset. Columns: transverse, coronal and sagittal images, respectively. (a): $CT_{original}$ and $CBCT_{original}$; (b): $CT_{deformed}^{demons}(\rightarrow CBCT_{original})$ and $CBCT_{original}$; (c): $CT_{deformed}^{DISC}(\rightarrow CBCT_{original})$ and $CBCT_{original}$; (d): $CT_{deformed}^{DISC}(\rightarrow CBCT_{original})$ and $CBCT_{corrected}$. The insets show the zoomed-in views.

On the other hand, the efficacy of CBCT intensity correction is demonstrated in Figure 4, where difference images between $CBCT_{original}^{primary}$ and $CBCT_{original}$, and between $CBCT_{original}^{primary}$ and $CBCT_{corrected}$ are shown. $CBCT_{original}^{primary}$ is considered as the CBCT image without any artifacts and thus with correct intensity. Before DIR, the scatter artifact is evident in CBCT$_{\text{original}}$ and hence a large deviation is observed from CBCT$_{\text{original}}^{\text{primary}}$ (Figure 4(a)). As DISC proceeds, the CBCT intensity is corrected gradually and the CBCT$_{\text{corrected}}$ is resulted, which has relatively small intensity inconsistency when compared with CBCT$_{\text{original}}^{\text{primary}}$ (Figure 4(b)).

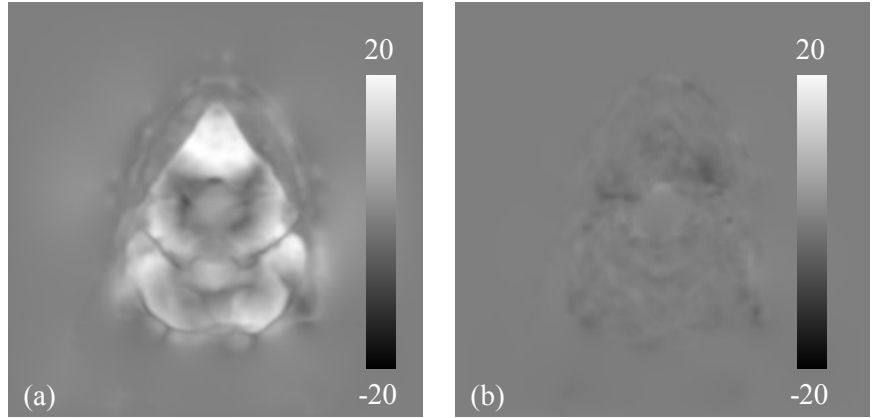

**Figure 3.** Difference of the deformation vector fields (display window: -20~20 voxels): (a): between CT$_{\text{deformed}}^{\text{demons}}(\rightarrow \text{CBCT}_{\text{original}})$ and CT$_{\text{deformed}}^{\text{demons}}(\rightarrow \text{CBCT}_{\text{original}}^{\text{primary}})$; (b): between CT$_{\text{deformed}}^{\text{DISC}}(\rightarrow \text{CBCT}_{\text{original}})$ and CT$_{\text{deformed}}^{\text{demons}}(\rightarrow \text{CBCT}_{\text{original}}^{\text{primary}})$.

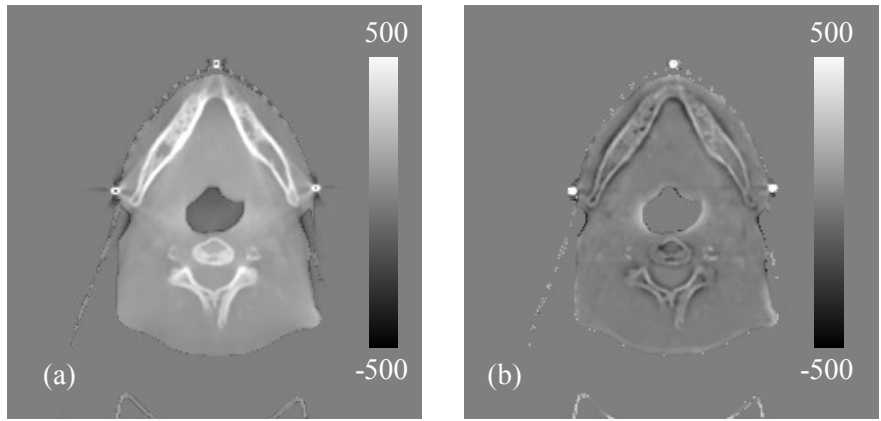

**Figure 4.** Difference images (display window: -500~500 HU): (a) between CBCT$_{\text{original}}^{\text{primary}}$ and CBCT$_{\text{original}}$; (b) between CBCT$_{\text{original}}^{\text{primary}}$ and CBCT$_{\text{corrected}}$.

*3.2 Clinical cases*

The performance of the DISC algorithm is further assessed on six head-and-neck cancer patient cases. An example result (Case 2) is shown in Figures 5 and 6. Misalignment is evident before registration (Figures 6 (a)-1~3). The results from the original demons algorithm are shown in Figure 5(c) and Figures 6(b)-1~3. As we can see, the anatomical structures in the deformed CT image are significantly distorted due to intensity inconsistency. This effect is more severe in regions that are dominated by artifacts (as indicated by arrows in Figure 5(c) and Figures 6(b)-1~3). The DISC algorithm (Figure 5(d) and Figures 6(c)-1~3), on the other hand, yields undistorted CT image that matches well with the original CBCT, which can also be seen from the comparison between $CT^{DISC}_{deformed}$ and $CBCT_{corrected}$ as shown in Figures 6(d)-1~3. This can be also verified by examining the histograms of $CT_{original}$, $CBCT_{original}$, $CT^{DISC}_{deformed}$, and $CBCT_{corrected}$ (Figure 7). We can see the remarkable histogram difference between $CT_{original}$ and $CBCT_{original}$ before DISC, and almost the same histogram distributions between $CT^{DISC}_{deformed}$ and $CBCT_{corrected}$. This indicates that the patch-based DISC algorithm can achieve a global match of intensity distribution, although the operations are purely local at each voxel.

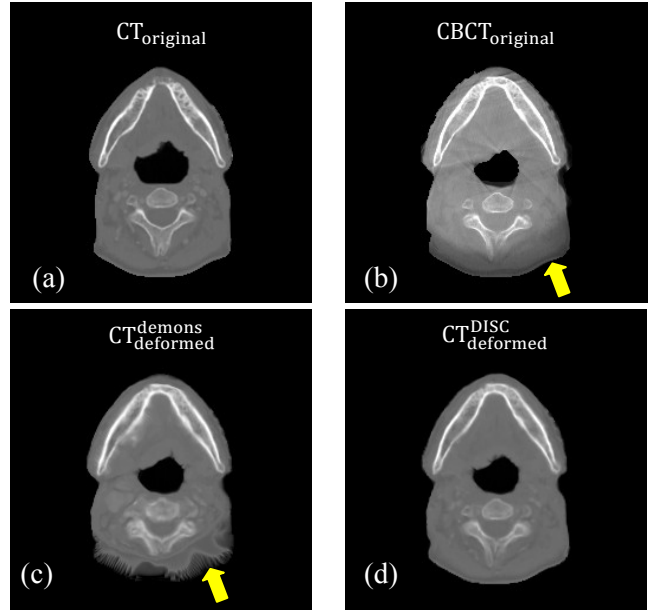

**Figure 5.** (a): $CT_{original}$; (b): $CBCT_{original}$; (c): $CT^{demons}_{deformed}$; (d): $CT^{DISC}_{deformed}$. The arrows indicate the region contaminated by artifacts (b) and the region with geometrical distortion (c).

In terms of DIR accuracy, the DISC algorithm is quantitatively evaluated using NMI, FSIM and Canny edge RMSE between $CT_{original}$ and $CBCT_{original}$, $CT^{demons}_{deformed}$ and $CBCT_{original}$, $CT^{DISC}_{deformed}$ and $CBCT_{original}$, and $CT^{DISC}_{deformed}$ and $CBCT_{corrected}$, as shown in Tables 1, 2 and 3. Both the original demons and DISC algorithms increase NMI (for all six cases) and reduce Canny edge RMSE (except for case 3 and 6 where the original demons algorithm generates even larger Canny edge RMSE) between CT and CBCT images. The FSIM for all six cases are increased after DISC. Interestingly, except for Case 2, the FSIM is decreased after DIR using the original demons algorithm. This is because the original demons algorithm distorts tissues significantly after DIR, which can be easily observed by human visual inspection but hard to be detected by NMI or Canny edge RMSE. It is much easier for FSIM to detect such changes, since it scores the similarity by mimicking how the human vision works. For all six cases, the average NMI increases from 0.62±0.02 to 0.63±0.02, the average FSIM increases from 0.91±0.04 to 0.94±0.02, and the average edge RMSE decreases from 0.24±0.03 to 0.21±0.03, when comparing DISC with the original demons algorithm.

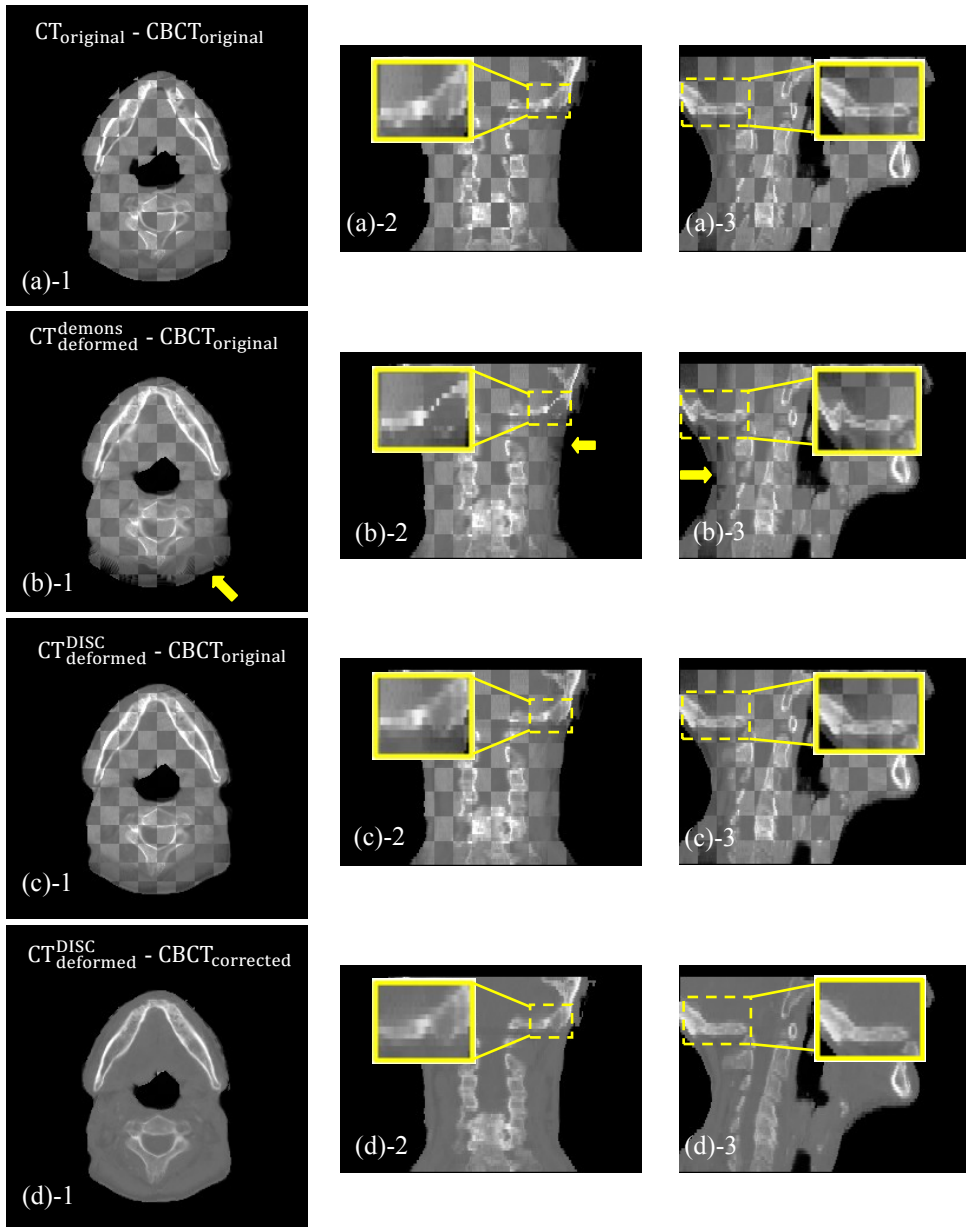

**Figure 6.** Checkerboard comparisons of one example clinical data. Columns: transverse, coronal and sagittal images, respectively. (a): $CT_{original}$ and $CBCT_{original}$; (b): $CT_{deformed}^{demons}$ and $CBCT_{original}$; (c): $CT_{deformed}^{DISC}$ and $CBCT_{original}$; (d): $CT_{deformed}^{DISC}$ and $CBCT_{corrected}$. The insets show the zoomed-in views, and the arrows indicate the regions contaminated by artifact or with geometrical distortion.

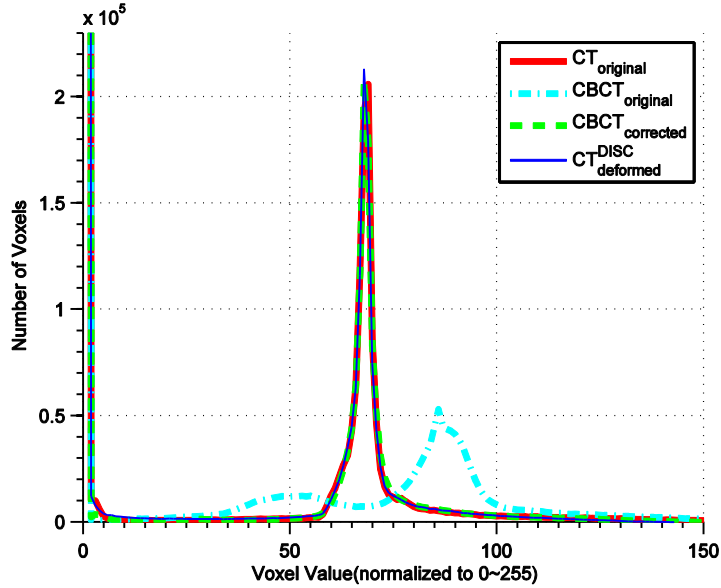

**Figure 7.** Histograms of $CT_{original}$, $CBCT_{original}$, $CT^{DISC}_{deformed}$ and $CBCT_{corrected}$.

**Table 1.** NMI between $CT_{original}$ and $CBCT_{original}$, $CT^{demons}_{deformed}$ and $CBCT_{original}$, $CT^{DISC}_{deformed}$ and $CBCT_{original}$, and $CT^{DISC}_{deformed}$ and $CBCT_{corrected}$.

|  | Case 1 | Case 2 | Case 3 | Case 4 | Case 5 | Case 6 |
|---|---|---|---|---|---|---|
| $CT_{original}$-$CBCT_{original}$ | 0.6034 | 0.6053 | 0.5972 | 0.6063 | 0.6481 | 0.6194 |
| $CT^{demons}_{deformed}$-$CBCT_{original}$ | 0.6089 | 0.6264 | 0.6032 | 0.6092 | 0.6566 | 0.6203 |
| $CT^{DISC}_{deformed}$-$CBCT_{original}$ | 0.6160 | 0.6266 | 0.6094 | 0.6191 | 0.6720 | 0.6302 |
| $CT^{DISC}_{deformed}$-$CBCT_{corrected}$ | 0.6811 | 0.6951 | 0.6939 | 0.6990 | 0.7454 | 0.6941 |

**Table 2.** FSIM and its standard deviation between $CT_{original}$ and $CBCT_{original}$, $CT^{demons}_{deformed}$ and $CBCT_{original}$, $CT^{DISC}_{deformed}$ and $CBCT_{original}$, and $CT^{DISC}_{deformed}$ and $CBCT_{corrected}$.

|  | Case 1 | Case 2 | Case 3 | Case 4 | Case 5 | Case 6 |
|---|---|---|---|---|---|---|
| $CT_{original}$-$CBCT_{original}$ | 0.9015 ±0.0153 | 0.9396 ±0.0147 | 0.9078 ±0.0162 | 0.9125 ±0.0071 | 0.9718 ±0.0026 | 0.9187 ±0.0132 |
| $CT^{demons}_{deformed}$-$CBCT_{original}$ | 0.8853 ±0.0144 | 0.9408 ±0.0191 | 0.8758 ±0.0158 | 0.8863 ±0.0257 | 0.9701 ±0.0026 | 0.8772 ±0.0135 |
| $CT^{DISC}_{deformed}$-$CBCT_{original}$ | 0.9106 ±0.0102 | 0.9543 ±0.0120 | 0.9301 ±0.0176 | 0.9314 ±0.0119 | 0.9797 ±0.0031 | 0.9227 ±0.0113 |
| $CT^{DISC}_{deformed}$-$CBCT_{corrected}$ | 0.9643 ±0.0066 | 0.9795 ±0.0124 | 0.9745 ±0.0156 | 0.9723 ±0.0088 | 0.9950 ±0.0022 | 0.9670 ±0.0119 |

**Table 3.** Canny edge RMSE between $CT_{original}$ and $CBCT_{original}$, $CT^{demons}_{deformed}$ and $CBCT_{original}$, $CT^{DISC}_{deformed}$ and $CBCT_{original}$, and $CT^{DISC}_{deformed}$ and $CBCT_{corrected}$.

|  | Case 1 | Case 2 | Case 3 | Case 4 | Case 5 | Case 6 |
|---|---|---|---|---|---|---|
| $CT_{original}$-$CBCT_{original}$ | 0.2707 | 0.2467 | 0.2623 | 0.2639 | 0.1863 | 0.2037 |
| $CT_{deformed}^{demons}$-$CBCT_{original}$ | 0.2541 | 0.2356 | 0.2764 | 0.2634 | 0.1752 | 0.2304 |
| $CT_{deformed}^{DISC}$-$CBCT_{original}$ | 0.2397 | 0.2165 | 0.2277 | 0.2257 | 0.1473 | 0.1979 |
| $CT_{deformed}^{DISC}$-$CBCT_{corrected}$ | 0.1923 | 0.1766 | 0.1855 | 0.1772 | 0.1014 | 0.1634 |

*3.3 Effect of patch size*

The path size in DISC has a considerable impact on the registration performance. Figures 8(a)-(l) show part of the $CT_{deformed}^{DISC}$ with different patch sizes in an example clinical case. This part of image is contaminated by artifacts most severely. As the patch size increases, the bones become distorted and the edges become blurred. When the patch size gets even larger, soft tissue region is distorted as well (Figures 8(g)-(l)). This effect is also observed in other clinical cases. The reason is that, as the patch size increases, it is more likely to have different structures included in the same patch, resulting in errors in intensity correction. Therefore, we use 3×3×3 patch size in this work for all simulation data and clinical data, and the patch size is kept constant at each image resolution level.

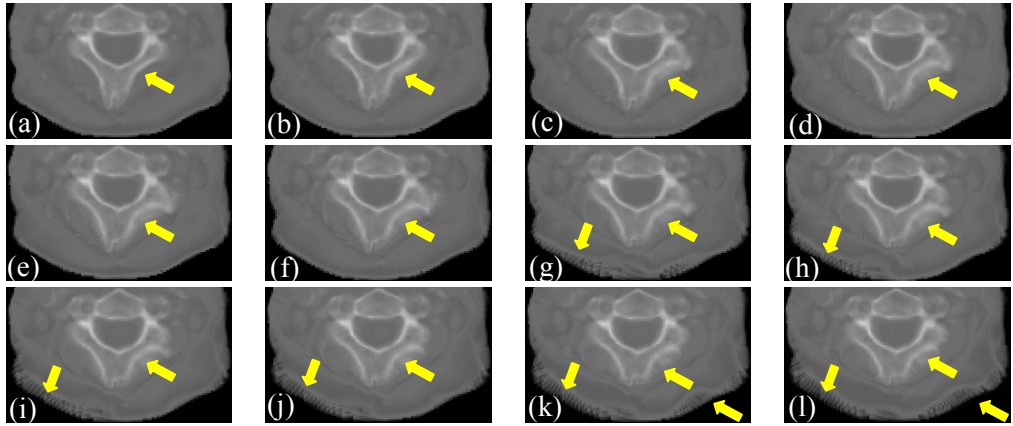

**Figure 8.** Effect of patch size on DISC performance. (a)-(l) are part of $CT_{deformed}^{DISC}$ obtained with patch size of 3×3×3, 5×5×3, 7×7×3, 9×9×3, 11×11×3, 13×13×5, 15×15×5, 17×17×5, 19×19×5, 21×21×5, 23×23×7, 25×25×7. The arrows indicate the regions of deterioration as the patch size increase.

*3.4 Variable a and b*

Figures 9(a) and 9(b) show one transversal slice of the distribution of the parameters $a$ and $b$ at the last iteration of DISC. Figures 9(c) and 9(d) are the patch standard deviation in $I_m$ and $I_s$ at the last iteration of DISC. In fact, $a$ is the ratio of the pixel value in Figure 9(c) to that in Figure 9(d) (see Equation (5)), and it controls the slope of the intensity transformation in each voxel. The value of $b$ represents a shift of the mean intensity value in a patch in $I_s$ to match that in $I_m$. We can see that the region with relatively high values of $b$ indicated by an arrow in Figure 9(b), showing a large shift is applied to change the mean intensity value of the patch, corresponds to the region where the large artifacts present in the original CBCT image. Figures 9(e) and 9(f) depict the evolution of the average value of *a* and *b* during the DISC iteration. We can see that both *a* and *b* converge nicely at both image resolution levels.

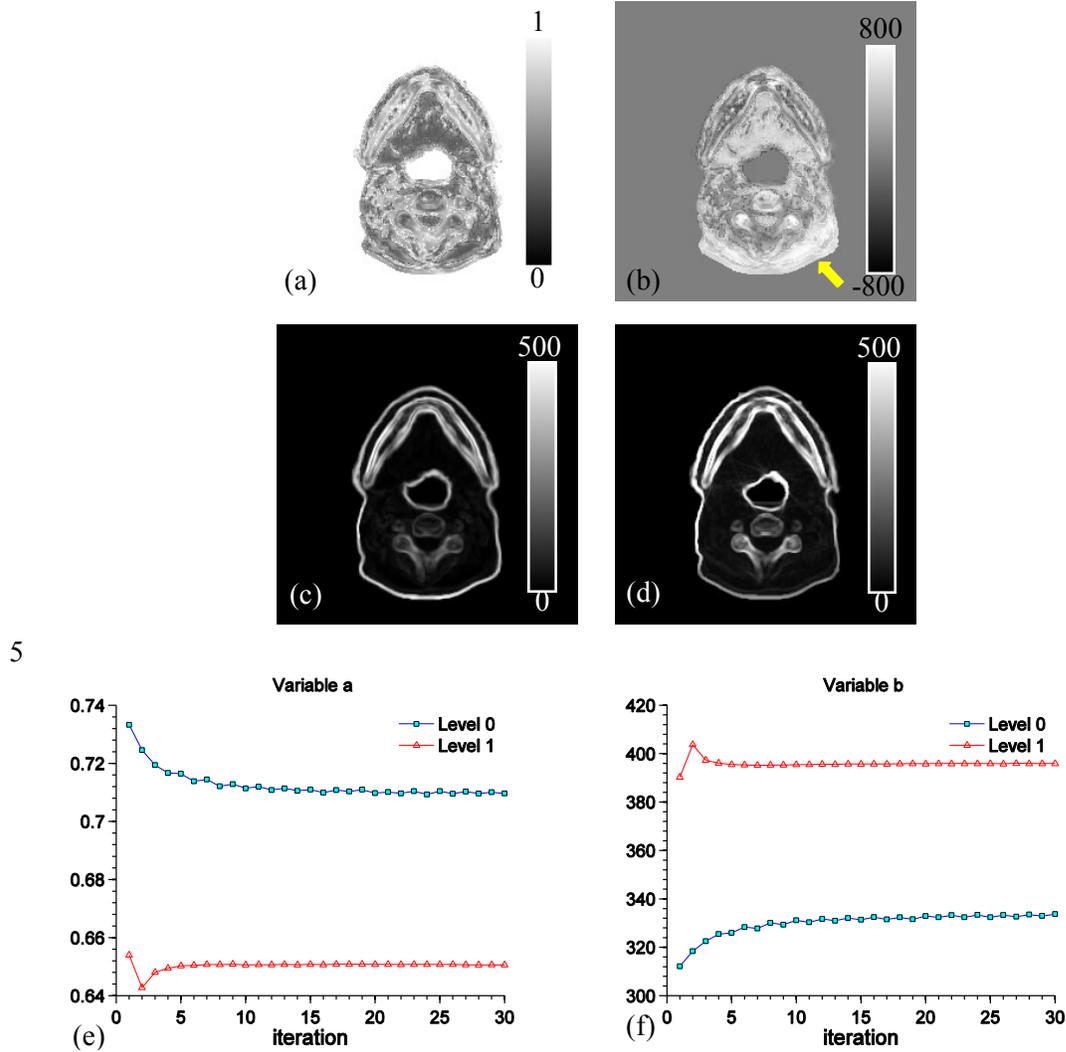

**Figure 9.** (a) and (b): Distributions of parameters *a* and *b* on one transversal slice, respectively. The arrow indicates where the artifact presents in the original CBCT image; (c) and (d): Standard deviation of image intensities in each patch in $I_m$ and $I_s$ at the last iteration; (e) and (f): Evolution of average *a* and *b* values with iteration at two image resolution levels, respectively. The results are for Case 2.

3.5 *Computational efficiency*

All the experiments in this study were conducted on an NVIDIA Telsa C1060 card with a total number of 240 processors of 1.3 GHz. It is also equipped with 4 GB DDR3 memory, shared by all processors. In the CUDA implementation of intensity correction kernel in DISC, the computation corresponding to each voxel is handled by a thread, which loops over all voxels in the patch centered at the voxel. Consequently, the computational time is highly dependent on the patch size. The most time consuming step in the calculation is the iterative interpolation/extrapolation procedure to get *a* and *b* values for voxels with $\xi = 1$. Currently, there is no efficient way for sparse data interpolation. Compared to ~20s needed with the original demons algorithm, the DISC algorithm takes about 69s for a patch size of 3×3×3 and an image size of 256×256×68.

**4. Discussion and Conclusions**

A fast, accurate, and robust CT-CBCT DIR algorithm is a key step in ART. Though many methods had been studied specifically for inter-modality image registration, which might be applied directly to CT-CBCT DIR, it is still tempting to use a more accurate and efficient intensity-based algorithm, such as demons, to deal with the DIR problem for CT and CBCT images. Other than modifying the DIR similarity metric, which is not straightforward in demons, we presented in this paper a patch-based intensity correction method that is performed in conjunction with demons, namely DISC, and evaluated it comprehensively with simulated data and clinical patient data. By incorporating the intensity correction at every iteration step, the DISC algorithm can robustly estimate the spatial transformation while guaranteeing the intensity consistency between CT and CBCT image.

Most previous studies focused on CBCT images without severe artifacts, in which case simple histogram matching (Hou *et al.*, 2011) or polynomial transformation model (Nithiananthan *et al.*, 2011) might be sufficient to describe the intensity relationship between the CT and CBCT image. In our DISC algorithm, intensity correction is performed on a voxel-by-voxel basis and the correction parameters are estimated by considering the voxels nearby. It is advantageous to do so, because usually, obtaining an ideal mathematical model to describe the intensity relation is not easy, if not impossible, when considering the existence of severe artifacts in CBCT images. While estimating the intensity transformation locally, one is able to make use of the prior information from CT image, and correct the intensity in CBCT image voxel-wise. By this means, we try to equalize the intensities distribution (the sample moments) inside each small patch with its corresponding patch in the other image, which is physically sensible. Embedding this additional correction step into the original demons, the intensity consistency requirement can be guaranteed.

Another merit of the DISC method is that it can yield the intensity corrected CBCT image at the end of the DIR. This intensity corrected CBCT image can be regarded as a CBCT image produced by the primary radiation. Hence, the DISC method, potentially, can be applied to the area of CBCT scatter removal and hence other clinically relevant tasks, e.g. radiation dose calculation based on CBCT.

**Acknowledgements**

This work is supported in part by NIH (1R01CA154747-01), the University of California Lab Fees Research Program, the Master Research Agreement from Varian Medical

Systems, Inc., and the grants from the National Natural Science Foundation of China (No.30970866)

**Appendix. Derivation of Equations (5) and (6)**

Given two groups of points $\mathbf{M}\{m_1, m_2, \cdots, m_n\}$ and $\mathbf{S}\{s_1, s_2, \cdots, s_n\}$, it is our objective to match the first and the second moments after a linear transformation. Specifically, we seek for constants $a(x)$ and $b(x)$ for the intensity transformation

$$\mathbf{S}' = a(x)\mathbf{S} + b(x), \tag{A.1}$$

s.t.

$$E(\mathbf{M}) = E(\mathbf{S}'), \tag{A.2}$$

$$E(\mathbf{M}^2) = E(\mathbf{S}'^2). \tag{A.3}$$

Because these two equations lead to

$$\text{STD}(\mathbf{M}) = \sqrt{E(\mathbf{M}^2) - (E(\mathbf{M}))^2} = \sqrt{E(\mathbf{S}'^2) - (E(\mathbf{S}'))^2} = \text{STD}(\mathbf{S}'), \tag{A.4}$$

taking the standard deviations on both sides of (A.1) yields

$$\text{STD}(\mathbf{M}) = \text{STD}(a(x)\mathbf{S} + b(x)) = a(x)\text{STD}(\mathbf{S}) \tag{A.5}$$

Hence

$$a(x) = \text{STD}(\mathbf{M})/\text{STD}(\mathbf{S}), \tag{A.6}$$

$$b(x) = E(\mathbf{M}) - a(x)E(\mathbf{S}). \tag{A.7}$$

**References**

Canny J 1986 A Computational Approach to Edge Detection *Pattern Analysis and Machine Intelligence, IEEE Transactions on* **PAMI-8** 679-98

Feldkamp L A, Davis L C and Kress J W 1984 Practical cone-beam algorithm *J. Opt. Soc. Am. A* **1** 612-9

Giles W, Bowsher J, Li H and Yin F-F 2011 Crescent artifacts in cone-beam CT *Medical Physics* **38** 2116

Grimmer R and Kachelriess M 2011 Empirical binary tomography calibration (EBTC) for the precorrection of beam hardening and scatter for flat panel CT *Medical Physics* **38** 2233-40

Gu X, Pan H, Liang Y, Castillo R, Yang D, Choi D, Castillo E, Majumdar A, Guerrero T and Jiang S B 2010 Implementation and evaluation of various demons deformable image registration algorithms on a GPU *Physics in Medicine and Biology* **55** 207-19

Guimond A, Roche A, Ayache N and Meunier J 2001 Three-dimensional multimodal brain warping using the demons algorithm and adaptive intensity corrections *IEEE Trans Med Imaging* **20** 58-69

Hou J, Guerrero M, Chen W and D'Souza W D 2011 Deformable planning CT to cone-beam CT image registration in head-and-neck cancer *Medical Physics* **38** 2088

Hsieh J, Molthen R C, Dawson C A and Johnson R H 2000 An iterative approach to the beam hardening correction in cone beam CT *Medical Physics* **27** 23-9

Jia X, Yan H, Cervino L, Folkerts M and Jiang S B 2012 A GPU Tool for Efficient, Accurate, and Realistic Simulation of Cone Beam CT Projections *submitted to Med Phys*

Lewis J H, Li R, Jia X, Watkins W T, Lou Y, Song W Y and Jiang S B 2011 Mitigation of motion artifacts in CBCT of lung tumors based on tracked tumor motion during CBCT acquisition *Physics in Medicine and Biology* **56** 5485-502

Li T, Schreibmann E, Yang Y and Xing L 2006 Motion correction for improved target localization with on-board cone-beam computed tomography *Physics in Medicine and Biology* **51** 253-67

Lu H, Reyes M, Serijovic A, Weber S, Sakurai Y, Yamagata H and Cattin P C 2010 Multi-modal diffeomorphic demons registration based on point-wise mutual information. In: *Proceedings of the 2010 IEEE international conference on Biomedical imaging: from nano to Macro,* (Rotterdam, Netherlands: IEEE Press) pp 372-5

Maltz J S, Gangadharan B, Bose S, Hristov D H, Faddegon B A, Paidi A and Bani-Hashemi A R 2008 Algorithm for X-ray scatter, beam-hardening, and beam profile correction in diagnostic (kilovoltage) and treatment (megavoltage) cone beam CT *IEEE Trans Med Imaging* **27** 1791-810

Meyer M, Kalender W A and Kyriakou Y 2010 A fast and pragmatic approach for scatter correction in flat-detector CT using elliptic modeling and iterative optimization *Phys Med Biol* **55** 99-120

Modat M, Vercauteren T, Ridgway G R, Hawkes D J, Fox N C and Ourselin S 2010 Diffeomorphic demons using normalized mutual information,

evaluation on multimodal brain MR images. ed B M Dawant and D R Haynor (San Diego, California, USA: SPIE) pp 76232K-8

Nithiananthan S, Schafer S, Uneri A, Mirota D J, Stayman J W, Zbijewski W, Brock K K, Daly M J, Chan H, Irish J C and Siewerdsen J H 2011 Demons deformable registration of CT and cone-beam CT using an iterative intensity matching approach *Medical Physics* **38** 1785

Pennec X, Cachier P and Ayache N 1999 Understanding the "Demon's Algorithm": 3D Non-Rigid registration by Gradient Descent. In: *2nd int. conf. on medical image computing and computer-assisted intervention (MICCAI'99) LNSC,* pp 597-605

Poludniowski G, Evans P M, Hansen V N and Webb S 2009 An efficient Monte Carlo-based algorithm for scatter correction in keV cone-beam CT *Physics in Medicine and Biology* **54** 3847

Rinkel J, Gerfault L, Estève F and Dinten J M 2007 A new method for x-ray scatter correction: first assessment on a cone-beam CT experimental setup *Physics in Medicine and Biology* **52** 4633

Rogelj P and Kovacic S 2006 Symmetric image registration *Med Image Anal* **10** 484-93

Samant S S, Xia J, Muyan-Özçelik P and Owens J D 2008 High performance computing for deformable image registration: Towards a new paradigm in adaptive radiotherapy *Medical Physics* **35** 3546

Schulze R, Heil U, Gross D, Bruellmann D, Dranischnikow E, Schwanecke U and Schoemer E 2011 Artefacts in CBCT: a review *Dentomaxillofacial Radiology* **40** 265-73

Sharp G C, Kandasamy N, Singh H and Folkert M 2007 GPU-based streaming architectures for fast cone-beam CT image reconstruction and demons deformable registration *Physics in Medicine and Biology* **52** 5771-83

Siewerdsen J H, Daly M J, Bakhtiar B, Moseley D J, Richard S, Keller H and Jaffray D A 2006 A simple, direct method for x-ray scatter estimation and correction in digital radiography and cone-beam CT *Med Phys* **33** 187-97

Siewerdsen J H and Jaffray D A 2001 Cone-beam computed tomography with a flat-panel imager: Magnitude and effects of x-ray scatter *Medical Physics* **28** 220

Sun M, Nagy T, Virshup G, Partain L, Oelhafen M and Star-Lack J 2011 Correction for patient table-induced scattered radiation in cone-beam computed tomography (CBCT) *Medical Physics* **38** 2058-73

Thirion J P 1998 Image matching as a diffusion process: an analogy with Maxwell's demons *Med Image Anal* **2** 243-60

Wang H, Dong L, O'Daniel J, Mohan R, Garden A S, Ang K K, Kuban D A, Bonnen M, Chang J Y and Cheung R 2005 Validation of an accelerated 'demons' algorithm for deformable image registration in radiation therapy *Physics in Medicine and Biology* **50** 2887-905

Yan H, Cervino L, Jia X and Jiang S B 2012 A comprehensive study on the relationship between the image quality and imaging dose in low-dose cone beam CT *Physics in Medicine and Biology* **57** 2063-80

Yan H, Mou X, Tang S, Xu Q and Zankl M 2010 Projection correlation based view interpolation for cone beam CT: primary fluence restoration in scatter measurement with a moving beam stop array *Physics in Medicine and Biology* **55** 6353-75

Yang D, Li H, Low D A, Deasy J O and Naqa I E 2008 A fast inverse consistent deformable image registration method based on symmetric optical flow computation *Physics in Medicine and Biology* **53** 6143-65

Zhang L, Zhang L, Mou X and Zhang D 2011 FSIM: a feature similarity index for image quality assessment *IEEE Trans Image Process* **20** 2378-86

Zheng D, Ford J C, Lu J, Lazos D, Hugo G D, Pokhrel D, Zhang L and Williamson J F 2011 Bow-tie wobble artifact: Effect of source assembly motion on cone-beam CT *Medical Physics* **38** 2508-14

Zhu L, Wang J and Xing L 2009a Noise suppression in scatter correction for cone-beam CT *Med Phys* **36** 741-52

Zhu L, Xie Y, Wang J and Xing L 2009b Scatter correction for cone-beam CT in radiation therapy *Medical Physics* **36** 2258